\documentclass[showpacs,preprintnumbers,amsmath,amssymb,prl,superscriptaddress,twocolumn,longbibliography,footinbib]{revtex4-1}

	\usepackage{natbib}
	\usepackage{txfonts}
	\usepackage{graphicx}
	\usepackage{dcolumn}
	\usepackage{bm}
	\usepackage{hyperref}
	\usepackage{color}
	\usepackage[table,xcdraw]{xcolor}
	\usepackage{amssymb}
	\usepackage{amsmath}
	\usepackage{float}
	\usepackage[justification=justified]{caption}
	\usepackage{epsfig}
	\usepackage{blindtext}

	\usepackage[utf8]{inputenc}

	\usepackage{silence}
	\newcommand{\beq}{\begin{equation}}
	\newcommand{\eeq}{\end{equation}}
	\newcommand{\bea}{\begin{eqnarray}}
	\newcommand{\eea}{\end{eqnarray}}

	\newcommand{\COSS}{Co$_3$Sn$_2$S$_2$}
	
	\newcommand{\kn}{{n\mathbf{k}}}

\begin{document}
	\title{Creating Weyl nodes and controlling their energy by magnetization rotation} 

	\author{Madhav Prasad Ghimire}
	\affiliation{Institute for Theoretical Solid State Physics, IFW Dresden, Helmholtzstr.~20, 01069 Dresden, Germany}
	\affiliation{Central Department of Physics, Tribhuvan University, Kirtipur, 44613, Kathmandu, Nepal}
	\affiliation{Condensed Matter Physics Research Center, Butwal-11, Rupandehi, Nepal}

	\author{Jorge I. Facio}
	\affiliation{Institute for Theoretical Solid State Physics, IFW Dresden, Helmholtzstr.~20, 01069 Dresden, Germany}

	\author{Jhih-Shih You}
	\affiliation{Institute for Theoretical Solid State Physics, IFW Dresden, Helmholtzstr.~20, 01069 Dresden, Germany}

	\author{Linda Ye}
	\affiliation{Department of Physics, Massachusetts Institute of Technology, Cambridge, Massachusetts 02139, USA}

	\author{Joseph G. Checkelsky}
	\affiliation{Department of Physics, Massachusetts Institute of Technology, Cambridge, Massachusetts 02139, USA}

	\author{Shiang Fang}
	\affiliation{Department of Physics, Harvard University, Cambridge, Massachusetts 02138, USA}

	\author{Efthimios Kaxiras}
	\affiliation{Department of Physics, Harvard University, Cambridge, Massachusetts 02138, USA}
	\affiliation{John A. Paulson School of Engineering and Applied Sciences, Harvard University, Cambridge, Massachusetts 02138, USA}

	\author{Manuel Richter}
	\affiliation{Institute for Theoretical Solid State Physics, IFW Dresden, Helmholtzstr.~20, 01069 Dresden, Germany}
	\affiliation{Dresden Center for Computational Materials Science (DCMS), TU Dresden, 01069 Dresden, Germany}
	\author{Jeroen van den Brink}
	\affiliation{Institute for Theoretical Solid State Physics, IFW Dresden, Helmholtzstr.~20, 01069 Dresden, Germany}

	\begin{abstract}
	As they do not rely on the presence of any crystal symmetry, Weyl nodes are robust topological features of an electronic structure that can occur at any momentum and energy. Acting as sinks and sources of Berry curvature, Weyl nodes have been predicted to strongly affect the transverse electronic response, like in the anomalous Hall or Nernst effects. 
	However, to observe large anomalous effects the Weyl nodes need to be close to or at the Fermi-level, which implies the band structure must be tuned by an external parameter, e.g. chemical doping or pressure.  
	Here we show that in a ferromagnetic metal tuning of the Weyl node energy and momentum can be achieved by rotation of the magnetization.
	Taking Co$_3$Sn$_2$S$_2$ as an example, we use electronic structure calculations based on density-functional theory to show that not only new Weyl fermions can be created by canting the magnetization away from the easy axis, but also that the Weyl nodes can be driven exactly to the Fermi surface.
	We also show that the dynamics in energy and momentum of the Weyl nodes strongly affect the calculated anomalous Hall and Nernst conductivities.
	\end{abstract}

	\maketitle

	Materials hosting unconventional quasiparticles, such as Weyl semimetals, constitute a framework with potential for novel electronic devices. One of the grounds for such expectation is the possibility of enhancing the response to external fields by taking advantage of the topological properties of the electronic states. For a material to specifically host Weyl fermions the spin degeneracy of the electronic bands has to be removed by breaking either inversion or time-reversal symmetry  ($\Theta$). 
	Karplus and Luttinger \cite{PhysRev.95.1154} first noticed that in a $\Theta$-broken system the spin-orbit coupling can introduce in the manifold of Bloch states a left-right asymmetry  which in turn, in the presence of an electric field, causes a Hall current at zero magnetic field.
	This scattering-independent mechanism originates in the so-called anomalous velocity of the wave-packets, 
	which can be written in terms of the Berry curvature of the Bloch states in momentum space.  
	Weyl nodes are monopoles of Berry curvature which implies first, that they can only be created and annihilated in pairs of opposite monopole charge 
	and second, that wave-packets made out of Weyl fermions can have a large anomalous velocity. 
	As this velocity is perpendicular to the electric field, Weyl systems can exhibit enhanced \textit{transverse} electronic responses, as in the Hall or Nernst effects.

	This effect has been argued to be at work in different materials in which the anomalous velocity contribution intrinsic to the band-structure is at the heart of enhanced electric and thermoelectric performance both in the regime of linear \cite{nakatsuji2015large,suzuki2016large,nayak2016large,wang2018large,liu2018giant,ghimire2018large} as well as in nonlinear response \cite{wu2017giant,facio2018strongly}.
	Still, a central problem for optimizing Berry-curvature-based effects is the energy of the Weyl fermions which currently is not a controlled variable from a material design point of view.
	Indeed, as the only symmetry restriction is to break inversion or $\Theta$, Weyl nodes can occur at any momentum and energy~\cite{Armitage2018,annurev-conmatphys-031016-025225,annurev-conmatphys-031016-025458}. 
	Here we propose a strategy to tune the Weyl node energy that relies on the subtle interplay between the magnetism and the energy dispersion in $\Theta$-breaking Weyl semimetals. 
	We essentially build on the fact that in magnetic materials with large spin-orbit coupling the orientation of the magnetization ($\mathbf{m}$) can substantially affect the energy of the Bloch waves. 
	We will show that this effect is strong enough to even create new Weyl nodes close to the Fermi surface and, {\it vice versa} annihilate Weyl pairs.

	We focus on the half-metal \COSS, a ferromagnet recently found to be a Weyl semimetal and to display a large anomalous Hall effect (AHE) \cite{liu2018giant,wang2018large}.
	Through electronic structure calculations based on Density Functional Theory (DFT) we show that canting the magnetization away from the easy axis leads to large displacements of the Weyl points in energy and in momentum space such that, at specific orientations, Weyl fermions can be tuned arbitrarily close to the Fermi surface and even be placed \textit{exactly} at it.
	Having established a controlled way to systematically reduce the distance between the Fermi energy and the Weyl node energy, we analyze how this tuning affects the anomalous Hall conductivity (AHC) and its thermal counterpart, the anomalous Nernst conductivity (ANC).
	Extensive calculations as a function of the magnetization direction reveal that the ANC in particular is highly susceptible to the Weyl dynamics: it displays sharp peaks when asymmetric hole and electron pockets exist nearby the band crossing that reaches the Fermi energy. 
	Measurements of the ANC in the presence of an external magnetic field that cants the magnetization will thus be able to signal the predicted appearance and tuning of Weyl nodes very close to the Fermi surface.

\begin{figure}[t]\center
	\includegraphics[width=8.5cm,angle=0,keepaspectratio=true]{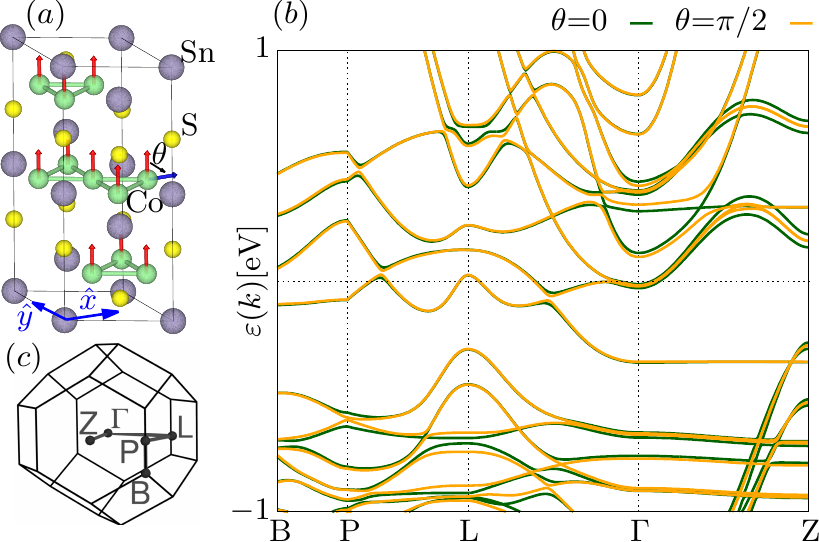}
		\caption{$(a)$ Crystal structure of \COSS. $(b)$ Band structure with the magnetization along the easy axis ($\theta=0$) or in the plane ($\theta=\pi/2$). The Fermi level at zero energy is indicated by a dotted line. $(c)$ Brillouin zone.
	}
	\label{bands}
	\end{figure}

	\textit{Controlling Weyl nodes with the magnetization. }
	\COSS\ is a layered ferromagnetic compound with Curie temperature $T_C=172\,$K \cite{schnelle2013ferromagnetic}. The crystal belongs to the space group $R$-$3m$ no.$ 166)$ with  lattice parameters $a=5.37\,$\AA\, and $c=13.18\,$\AA.
	The system consists of quasi-2D Co$_3$Sn layers separated by S$_2$Sn layers with the particularity that the stack of cobalt atoms forms a kagome lattice, see Fig. \ref{bands}$(a)$.
	The spin-orbit coupling induces a magnetic anisotropy which favours $\mathbf{m}$ to lie along the three-fold rotation axis $\hat{z}$ \cite{liu2018giant}. Recently, different groups \cite{liu2018giant,wang2018large} have studied the topological structure of the magnetic ground-state and have found the existence of six Weyl nodes lying 60$\,$meV above the Fermi energy. 
	We performed DFT calculations for different ferromagnetic configurations in which $\mathbf{m}$ lies in the $xz$ plane at an angle $\theta$ from $\hat{z}$.  
	\footnote{We used the all-electron full-potential local-orbital (\textsc{FPLO}) code \cite{koepernik1999full} (18.00-52) in the fully relativistic mode using the GGA approximation (PBE-96) \cite{perdew1996generalized} with an enhanced basis set \cite{lejaeghere2016reproducibility} and a linear tetrahedron method with $12\times12\times12$ intervals in the Brillouin zone.}
	Figure \ref{bands}$(b)$ presents the electronic band structure for the limit cases $\theta=0,\, \pi/2$. 
	The calculations indicate that the magnetic ground state corresponds to $\theta=0$, with a magnetic moment of 0.33 $\mu_B$, which agrees well with experimental results based on magnetization measurements, 0.29 $\mu_B$ \cite{vaqueiro2009powder} or 0.31 $\mu_B$ \cite{schnelle2013ferromagnetic}. 
	The calculated magnetic anisotropy energy is $\sim 0.19\,$meV per Co atom. 
	The field required to rotate the moment into the plane at zero temperature is estimated to be $\sim 26\,$Tesla. 
	\footnote{The magnetic anisotropy estimation is based on DFT total energies and on considering the leading term in the anisotropy energy $E=K_0\sin^2(\theta) - M H\sin(\theta)$ where $M=|\mathbf{m}|$ and $H$ is the external field \cite{chikazumi2009physics}. The result agrees well with magnetization measurements as a function of magnetic field which indicate that at 10\,K a field of 21.3 Tesla is required to make the in-plane magnetization reach the value presented by the out-of-plane component at zero field. We did not consider effects of the external field other than the rotation of $\mathbf{m}$.}

While for both magnetic configurations the system displays flat bands along $\Gamma$-$Z$, as observed in other layered kagome materials \cite{koudela2008magnetic,ochi2015robust,mazin2014theoretical,ye2018massive}, the rotation of $\mathbf{m}$ induces appreciable changes in the energy splitting between some of the bands, and even makes some of the original crossings to become avoided and {\it vice versa}. Indeed, the search for Weyl points in an energy range [-100,150]\,meV as implemented in FPLO~\cite{koepernik1999full,Koepernik2016,Lau2017} shows that their number increases from 6 to 26 as $\theta$ changes from 0 to $\pi/2$. 
\footnote{For the search of Weyl points and the calculation of $\Omega_{n,c}(\mathbf{k})$ we used the \textsc {PYFPLO} module of \textsc{FPLO}. For each angle $\theta $, we built a tight-binding Hamiltonian by projecting the Bloch states onto atomic-orbital-like Wannier functions associated with Co 3d and 4s states, Sn 5p and 5s and S 3p.}

The large change in the number of Weyl points caused by rotating $\mathbf{m}$ from out-of-plane to in-plane raises the question of how the system evolves between these configurations. Our calculations show that there are two regimes. The first regime occurs at small angles in which the main effect is the splitting in energy of the original six Weyl points due to the breaking of the C$_3$ symmetry. The second one is signalled by the creation of Weyl points and by rich dynamics in which Weyl fermions travel along large trajectories in momentum space.

	\begin{figure}[t]\center
	\includegraphics[width=8.5cm,angle=0,keepaspectratio=true]{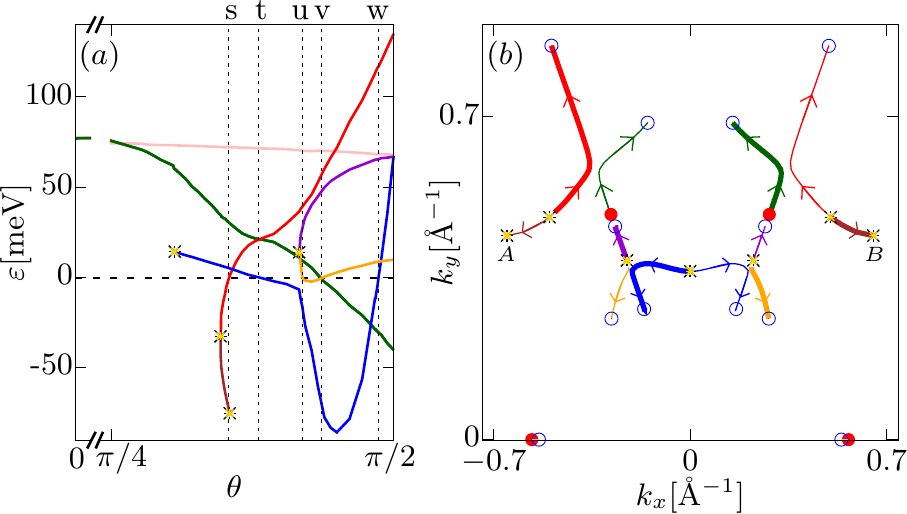}
		
	\vspace{1mm}
	\includegraphics[width=8.5cm,angle=0,keepaspectratio=true]{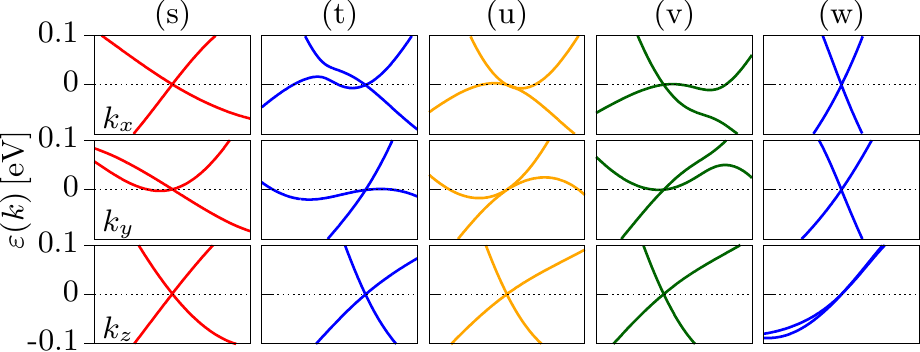}

		\caption{ Weyl nodes dynamics. 
	$(a)$ Energy of the Weyl nodes as a function of $\theta$. 
	Different colors correspond to sets of Weyl points of different energy. The yellow marks signal the creation or annihilation of Weyl nodes. 
	Vertical lines indicate angles (labeled s,t,...,w) at which Weyl fermions place at the Fermi surface. 
	$(b)$ Projection of Weyl node trajectories on the semiplane $k_z=0$ with $k_y\geq 0$. 
		Filled-red and empty-blue circles correspond to Weyl points at $\theta=0$ and $\theta=\pi/2$, respectively. 
	The thin (thick) lines indicate the trajectories of Weyl points of positive (negative) chirality and follow the color code of $(a)$ while the arrows indicate the sense of motion as $\theta$ increases.
	Weyl points $A$ and $B$ annihilate each other by approaching the Brillouin zone border when their separation reaches a reciprocal lattice vector.
	(s-w) For the angles in which a Weyl node crosses the chemical potential, energy dispersion along paths centered at the corresponding Weyl point. The paths have lenght $2\pi/(5a)$ and the rows from top to bottom correspond to the dispersion along $k_x$, $k_y$ and $k_z$, respectively.
	}
	\label{energy}
	\end{figure}
	We first analyze the constraints imposed by the symmetry on the evolution of the Weyl nodes as a function of the magnetization direction. The non-magnetic point symmetry group is $D_{3d}(b)$ \footnote{See Supplementary materials for $(i)$ Flux of Berry curvature for different orientations of $\mathbf{m}$, $(ii)$ Projection of Weyl-node trajectories in the full Brillouin zone and $(iii)$ Derivation of Eq. (3) and further details about the numerical calculations.}.
A finite $\mathbf{m}$ breaks $\Theta$ and affects some of the unitary symmetry operations. 
For instance, for $\mathbf{m}\parallel\hat{z}$, the mirror symmetries $M$ associated with planes that contain $\hat{z}$ are broken while the system remains invariant under the operation $\Theta \cdot M$. The same occurs with the two-fold rotations of axes perpendicular to $\hat{z}$. 
As a result, for such magnetic configuration, there are twelve point symmetry operations and the application of these to a Weyl node that lies in the $k_y=0$ plane leads to a total of six Weyl nodes. 
Upon canting $\mathbf{m}$ at an angle $\theta$ towards $\hat{x}$, the only remaining symmetries -- in addition to the identity and to inversion -- are $\Theta \cdot C_2(y)$ and $\Theta \cdot M(y)$. This forces the six Weyl nodes to split in energy as a doublet (formed by those with $k_y=0$) and a quartet.  
Fig. \ref{energy}$(a)$ shows the calculated energy of the Weyl nodes as a function of $\theta$. While for $\theta<\pi/4$ the splitting between the doublet and the quartet is smaller than 4\,meV, further increasing $\theta$ shifts the energy of the quartet from \textit{above} to \textit{below} the Fermi level. 
The vertical lines in Fig. \ref{energy}$(a)$ indicate the angles (labeled s,t,..,w) at which a Weyl node crosses the Fermi surface. 
As shown in Fig. \ref{energy}(s-w), the energy dispersion close to the Weyl points reaching the Fermi energy can differ significanly between the different cases \footnote{Close to $\theta$$=$v there are actually two sets of Weyl points reaching the Fermi surface. In Fig. \ref{energy}(v) for simplicity we chose to show the dispersion of only one of these sets. Also, we omit in Fig. \ref{energy} four Weyl nodes in the plane $k_y=0$ that are stable only in a small angle range close to $\pi/2$.}.

	A detailed inspection of the Weyl node trajectories indicates that the remarkably large change in energy goes hand in hand with a large displacement in momentum space. 
	Fig. \ref{energy}$(b)$ shows the projection of the computed trajectories on the semiplane $k_z=0$ with $k_y \geq 0$, illustrating that $\theta$ controls a rich dynamics that includes the creation and annihilation of Weyl nodes. 

	\textit{Experimental consequences of the Weyl dynamics. }
	We analyze next observables that can make evident the rich Weyl dynamics induced by rotating $\mathbf{m}$. 
	We focus on transverse response functions to an electric field $\mathcal{E}$, a natural choice since Weyl nodes are sources or sinks of Berry curvature and therefore affect the carrier velocity through the anomalous velocity \cite{PhysRevB.81.052303}.
	This causes the AHE, a Hall response $j_a = \sigma_{ab} \mathcal{E}_b$ in a $\Theta$-broken state without external magnetic field, where $\sigma_{ab}$ is 
	\begin{equation}
		\sigma_{ab} = \frac{-e^2}{\hbar}\int \frac{d^3\mathbf{k}}{(2\pi)^3} \sum_{n,c} \epsilon_{abc} \Omega_{n,c}(\mathbf{k}) f_0(E_{n\mathbf{k}}).
	\label{eq:hall}
	\end{equation}
	Here, $\epsilon_{abc}$ is the Levi-Civita tensor, $E_\kn$ the energy dispersion of the $n$-th band, $f_0$ the equilibrium Fermi distribution and $\Omega_{n,c}$ is the $c$ component of the Berry curvature, $\mathbf{\Omega}_n(\mathbf{k})=\mathbf{\nabla}_\mathbf{k}\times\mathbf{A}_n(\mathbf{k})$, where $\mathbf{A}_n(\mathbf{k})=-\mathrm{i}\langle u_\kn | \partial_\mathbf{k}  u_\kn \rangle$ \cite{PhysRevB.76.195109}. 
	At temperature $T$, the anomalous velocity can also generate a transverse flow of entropy resulting in an anomalous Nernst effect (ANE): a temperature gradient generates a Hall current, $j_a =-\alpha_{ab} \mathcal{\nabla}_b T$, \cite{PhysRevLett.97.026603,PhysRevB.78.174508,PhysRevB.90.165115,PhysRevB.93.035116,guo2014anomalous,PhysRevB.96.115202,PhysRevB.96.235116,saha2018anomalous,doi:10.1063/1.5037551,doi:10.1146/annurev-conmatphys-033117-054129}, where 
	\begin{equation}
		\alpha_{ab} = \int \frac{d^3\mathbf{k}}{(2\pi)^3} \sum_{n,c} \epsilon_{abc} \Omega_{n,c}(\mathbf{k}) s(E_{n\mathbf{k}}).
	\label{eq:nernst}
	\end{equation}
	Here, $s(E_{n\mathbf{k}})= f_0(E_{n\mathbf{k}})(E_{n\mathbf{k}}-\mu)/T -k_B \log\Big(1-f_0(E_{n\mathbf{k}}) \Big)$ is the entropy density and $\mu$ the chemical potential. 
	At zero temperature the entropy vanishes and so does $\alpha_{ab}$, while at finite temperature 
	$s(E_{n\mathbf{k}})$ is maximum at the chemical potential and decreases exponentially away from the Fermi surface. 
	Via joint measurement of electric and thermoelectric coefficients one can determine both these tensors \cite{PhysRevLett.119.056601}. 

	\begin{figure}[t]\center
	\includegraphics[width=8.5cm,angle=0,keepaspectratio=true]{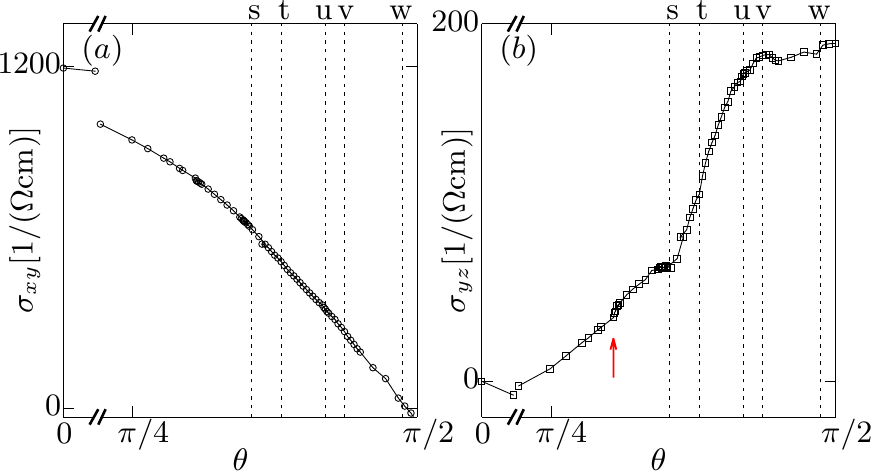}

		\vspace{1mm}
	\includegraphics[width=8.5cm,angle=0,keepaspectratio=true]{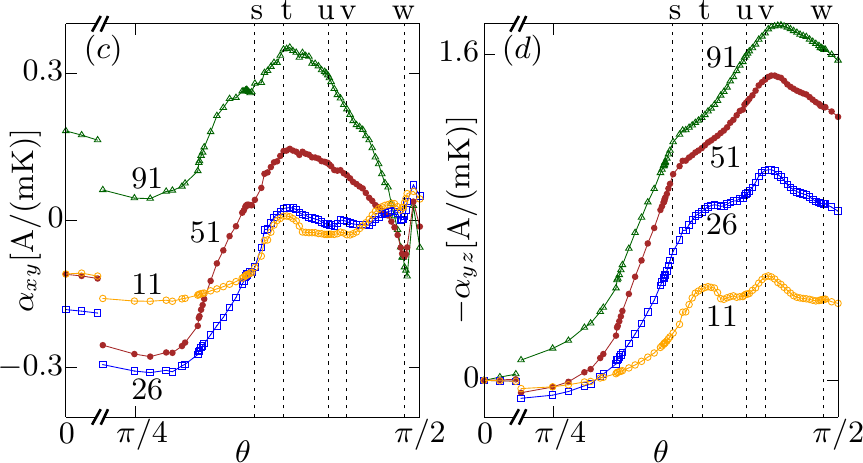}
		\caption{$(a-b):$ Anomalous Hall conductivity ($xy$ and $yz$ components) at zero temperature as a function of $\theta$. 
		The red arrow indicates an angle at which pairs of Weyl nodes are created. 
		$(c-d)$ Nernst conductivity as a function of $\theta$ for different temperatures (indicated in units of K for each curve). Note that $\alpha_{yz}$ is multiplied with -1. 
	}
	\label{anomalous}
	\end{figure}

Figure \ref{anomalous}$(a)$-$(b)$ shows the calculated AHCs as a function of $\theta$ 
\footnote{For numerical integration of Eq. \ref{eq:hall}, we used a mesh of density $400^3/V_{BZ}$ where $V_{BZ}$ is the Brillouin zone volume. The results of entropy-flow density were obtained with a density of 1500/$V_{BZ}$ points and considering states in the energy windows [-20,20]\,meV (see Supp. Materials).}. 
	For $\theta$$=$$0$, only $\sigma_{xy}$ is non-zero and its very large value $1200\,(\Omega$cm$)^{-1}$ is in good agreement with values reported of 1310 \,$(\Omega$cm$)^{-1}$ \cite{wang2018large} and 1100 $\,(\Omega$cm$)^{-1}$\cite{liu2018giant}. When $\mathbf{m}$ acquires a finite projection on $\hat{x}$, the only symmetry involving a mirror plane is $\Theta \cdot M(y)$ and both $\sigma_{xy}$ and $\sigma_{yz}$ are symmetry-allowed. 
	Namely, under mirror-symmetry $\mathbf{\Omega}$ behaves as a pseudo-vector so that $M(y)\Omega_y(\mathbf{k})=\Omega_y(k_x,-k_y,k_z)$ while $M(y)\Omega_\parallel(\mathbf{k})=-\Omega_\parallel(k_x,-k_y,k_z)$, where $\Omega_\parallel$ refers to the components of $\mathbf{\Omega}$ along $\hat{x}$ and $\hat{z}$ (parallel to the mirror plane). Therefore, if $\Theta \cdot M(y)$ is a symmetry, the Berry curvature satisfies $\Omega_y(\mathbf{k})=-\Omega_y(-k_x,k_y,-k_z)$ and $\Omega_\parallel(\mathbf{k})=\Omega_\parallel(-k_x,k_y,-k_z)$ and these constraints only make $\sigma_{xz}$ vanish.

	As $\theta$ increases, $\sigma_{xy}$ and $\sigma_{yz}$ follow opposite trends and at $\theta=\pi/2$, $\sigma_{yz}\sim200\,(\Omega$cm$)^{-1}$ is much larger than  $\sigma_{xy}$. At this angle, the $xy$ component does {\it not} vanish as it is still symmetry-allowed. The smallness of its value is still related to the effects of the symmetries on the Berry curvature flux. 
In this compound, the large AHC arises mainly from the nodal lines that become gapped when spin-orbit is included.
When $\mathbf{m}$ points along $\hat{z}$, the combination of $C_3$ and the mirror planes $\Theta\cdot M(x)$, $\Theta\cdot M(y)$ forces the different patches of the nodal lines to contribute \textit{additively} to the flux of $\Omega_z$ (see Supplementary Materials \cite{Note4}). 
	When $\mathbf{m}$ acquires a component along $\hat{x}$, the reduction in the symmetries removes this constraint and large cancelations occur due to the opposite flux of $\Omega_z$ at different points $(k_x,k_y)$. 
	This geometrical effect controls the overall evolution of $\sigma_{xy}$ which is, therefore, quite smooth and monotonous. 
	The $yz$ component lacks this geometrical effect and, as a result, is more sensitive to detailed changes in the band-structure, in particular to the Weyl dynamics, as shown in Figure \ref{anomalous}$(b)$.  
	For instance, the red arrow in the plot indicates an angle at which the creation of Weyl nodes leads to a step-like increase of  $\sigma_{yz}$. 
	As $\theta$ increases further, a set of Weyl nodes reaches the Fermi energy at $\theta$$=$s and $\sigma_{yz}$ exhibits a small plateau. The energy dispersion of these Weyl fermions is shown in Figure \ref{energy}$($s$)$ and makes clear that this node is type-I and therefore the observed plateau is consistent with the general prediction for such Weyl nodes when their energy approaches the Fermi level \cite{PhysRevLett.113.187202}.

Even though AHC measurements will be a useful tool to observe experimentally the predicted Weyl dynamics, the resulting changes related to the Weyl nodes are not large, providing significant experimental challenges. It turns out that the ANE shows much larger changes when Weyl nodes approach the Fermi energy. 
Figures \ref{anomalous}$(c)$-$(d)$ show the ANCs as a function of $\theta$ for different values of $T$. 
Since thermal fluctuations of the Co magnetic moments are not considered in Eq. (\ref{eq:nernst}), we restrict our analysis to a range of $T$ where the magnetization is essentially saturated ($T<100$\,K)\cite{schnelle2013ferromagnetic}. 
Remarkably, the dependence on $\theta$ is non-monotonous and includes sharp peaks at specific angles. The peaks are narrower for lower $T$ and are centered at some of the angles in which a Weyl node crosses the Fermi energy (specifically, $\theta$$=$t and v).
Our calculations show that even if the features become broader as $T$ increases the non-monotonous features remain clearly visible. The magnitude of the enhacements is worth noting: at 91\,K, starting from $\sim0.1\,$A(mK)$^{-1}$ at $\pi/4$, $|\alpha_{yz}|$ reaches a maximum $1.8\,$A(mK)$^{-1}$. 
This value is five times the maximum obtained in Mn$_3$Sn \cite{PhysRevLett.119.056601,ikhlas2017large} and half of the giant ANC in Co$_2$MnGa \cite{sakai2018giant}.

\textit{Entropy flow and Weyl fermions. } 
To understand why and when a peak in the ANC occurs, it proves useful to analyze how different carriers contribute to the ANC. 
Since in a fermionic system the available entropy is restricted to states of energy $\sim$$k_B T$ around $\mu$, we are interested in resolving the contribution of different electronic states within this thermal energy windows. 
We recast Eq. (\ref{eq:nernst}) as 
\begin{equation}
\alpha_{ab} = -\frac{1}{e} \int_{-\infty}^\infty d\varepsilon \frac{\partial\sigma_{ab}}{\partial \varepsilon} s(\varepsilon),
\label{eq:nen}
\end{equation}
(see Supplementary Materials \cite{Note4}).
This equation has the overall shape of the Mott relation --involving the derivative with respect to energy of the AHC and the entropy-- but holds for any temperature and motivates us to define the entropy-flow density per energy, $s_{f,ab}(\varepsilon) = \frac{\partial\sigma_{ab}}{\partial \varepsilon} s(\varepsilon)$.
The entropy flow $s_{f,ab}(\varepsilon)$ measures the contribution of carriers of different energy to the ANC and, therefore, its calculation allows us to assess to what extent different sets of Weyl points contribute to the ANC.

\begin{figure}[t]\center
\includegraphics[width=8.5cm,angle=0,keepaspectratio=true]{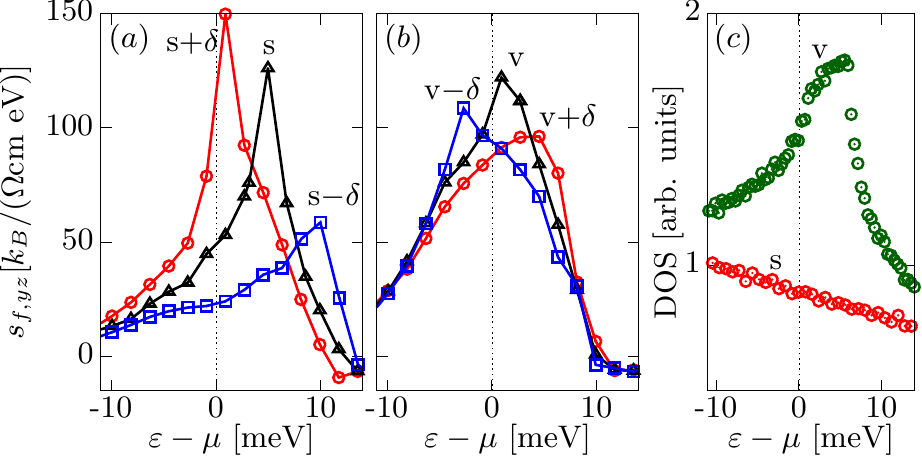}
	\caption{Transverse entropy flow. $(a)$ Entropy-flow density as a function of energy for $\theta$=s, s$\pm\delta$ with $\delta\sim4^\circ$. $(b)$ Entropy-flow density for $\theta=$v,v$\pm\delta$. The data correspond to the $yz$ component and at $T=51$\,K. $(c)$ DOS as a function of energy associated with the states close to the corresponding Weyl nodes at the Fermi surface for $\theta$$=$s and v.} 
\label{energy_der}
\end{figure}
We first analyze $\theta$$=$s. We focus on the $yz$ component which is the one displaying larger changes.
Fig. \ref{energy_der}$(a)$ presents $s_{f,yz} (\varepsilon)$ for $\theta$$=$s and for two other angles close to it, s$\pm$$\delta$ with $\delta$$\sim$$4^\circ$.
For each angle, the entropy flow presents a peak as a function of energy which highlights the energy of the carriers that contribute the most to the ANC.
Notice that this peak needs not to be centered at the energy of maximum entropy ($\varepsilon=\mu$). 
For instance, at $\theta$$=$s$-$$\delta$ the largest contribution is found $\sim$10\,meV above $\mu$. As $\theta$ increases, this peak achieves a larger value and its position shifts to smaller energy in a rather monotonous way.
In particular, at $\theta$$=$s the contribution of carriers at the Fermi surface is markedly smaller than that of the states conforming the entropy-flow peak. 
This explicitly shows that the set of Weyl nodes crossing the Fermi energy at this angle does not contribute significantly to the ANC.

The situation is different at angles $\theta$$=$t and v.
Fig. \ref{energy_der}$(b)$ shows $s_{f,yz}(\varepsilon)$ for $\theta$$=$v and $\theta$$=$v$\pm$$\delta$ and makes clear that the entropy-flow peak places at the chemical potential when the Weyl nodes reach the Fermi surface ($\theta$$=$v). This indicates that in this case the Weyl points indeed dominate the ANC.

The different contribution of different Weyl nodes to the ANC can be traced to the band structure close to the Weyl node in each case [see Fig. \ref{energy}(s-t-v)].
While at $\theta$$=$s the Weyl point at the Fermi surface is type-I with only a small tilt, at angles $\theta$$=$t and v the dispersion of one of the bands involved in the crossing at the Fermi level is such that it produces asymmetric electron and hole pockets close to the node. 
To illustrate this, Fig. \ref{energy_der}$(c)$ shows that the density of states (DOS) in a sphere of radius $\pi/(5 a)$ centered at the corresponding Weyl node at the Fermi surface is larger and more electron-hole asymmetric for $\theta$$=$v than for s. 
In an analagous way, for $\theta$$=$v, the more asymmetric distribution of states close to the nodes at the Fermi surface contributes to a larger  ${\partial \sigma_{yz}}/{\partial \varepsilon}$ and hence a larger entropy flow.
We thus associate the enhanced ANC at some of the angles at which a Weyl node reaches the Fermi surface with having asymmetric hole and electron pockets close to the Weyl point at the Fermi energy. 

In summary, we have shown that a rich dynamics of Weyl points results from the canting of the magnetization in the half-metallic ferromagnet \COSS: Weyl nodes are created, annihilated and shifted in energy-momentum space over large distances. 
This dynamics can be used to place Weyl fermions exactly at the Fermi surface, which is reflected in the calculated anomalous Hall conductivity and leads to sharp peaks in the anomalous Nernst conductivity that survive up to relatively high temperatures. Whereas we established the effect of rotation of the magnetization on Weyl nodes in detail in \COSS, our symmetry considerations suggest that the Weyl node energy and position in momentum space can be finely tuned in time-reversal-broken Weyl semimetals in general, including ferrimagnetic ones.
This provides both a strong motivation and a pathway to experimentally probe, manipulate and control Weyl-fermion transport properties via external magnetic fields in \COSS\, and other magnetic Weyl semimetals.

\textit{Acknowledgments. }
M.P.G. and J.I.F. contribute equally to this work.
We thank Ulrike Nitzsche for technical assistance.  M.R and J.v.d.B. acknowledge support from the German Research Foundation (DFG) via SFB 1143, project A5.
M.P.G. thanks the Alexander von Humboldt Foundation for financial support through the Georg Forster Research Fellowship Program.
J.I.F. and J.-S.Y. thank the IFW excellence programme.
J.G.C. acknowledges support by the Betty Moore Foundation EPiQS Initiative, grant GBMF3848.
L.Y. acknowledges support by the Tsinghua Education Foundation.
E.K. and S.F. were supported by the STC Center for Integrated Quantum Materials, NSF Grant No. DMR-1231319 and by ARO MURI Award No.W911NF-14-0247.

\onecolumngrid
\section{Supplementary material}
\subsection{Flux of Berry curvature for different orientations of the magnetization}

Table S1 lists the point symmetries of the material for different magnetic configurations.
To illustrate how the reduction of symmetries affects the anomalous Hall conductivity, here we present the flux of $\Omega_z$ integrated over $k_z$: 
$\phi_{\Omega z}(k_x,k_y) = \sum\limits_n \int dk_z \Omega_{n,z}(k_x,k_y,k_z) f_0(E_{n,k})$, where the integration is performed in the primitive Brillouin zone.
Figure S1$(a)$ shows $\phi_{\Omega z}$ computed with the magnetization parallel to $\hat{z}$.
It is noticeable that all the hot spots contribute with the same sign to $\sigma_{xy}$.
Symmetries play a role in this, for instance, enforcing the large contributions observed nearby to the six Weyl points to have the same sign.
Namely, Weyl points related by $C_3$ rotations 
--sets $\{A, D, E\}$ and  $\{B, C, F\}$ in Figure S1$(a)$--
are forced to contribute in the same amount to the flux of $\Omega_z$. 
Additionally, $\Theta \cdot M$, with $M=M(x), M(y) \text{ and } M(xy)$, makes the Berry curvature flux generated by these two sets to also contribute with the same sign since. For example, when $\Theta \cdot M(y)$ is a symmetry one has  $\Omega_z(k_x,k_y,k_z)=\Omega_z(-k_x,k_y,-k_z)$, and therefore, $\phi_{\Omega z}(\vec{k}_C) = \phi_{\Omega z}(\vec{k}_E)$.
When $\mathbf{m}$ has a finite projection along $\hat{x}$, $C_3$ is broken enabling cancellations in the total flux, as shown in Figure S1$(b)$.

\begin{table}[h]
\begin{tabular}{|
>{\columncolor[HTML]{FFFFFF}}c |
>{\columncolor[HTML]{FFFFFF}}c |
>{\columncolor[HTML]{FFFFFF}}c |
>{\columncolor[HTML]{FFFFFF}}c |
>{\columncolor[HTML]{FFFFFF}}c |
>{\columncolor[HTML]{FFFFFF}}c 
>{\columncolor[HTML]{FFFFFF}}c 
>{\columncolor[HTML]{FFFFFF}}c 
>{\columncolor[HTML]{FFFFFF}}c 
>{\columncolor[HTML]{FFFFFF}}c 
>{\columncolor[HTML]{FFFFFF}}c 
>{\columncolor[HTML]{FFFFFF}}c 
>{\columncolor[HTML]{FFFFFF}}c }
\hline
Non-magnetic                   & $E$ & $I$ & $C_3$-$(z)$         & $C_3(z)$          & \multicolumn{1}{c|}{\cellcolor[HTML]{FFFFFF}$C_2(xy)$}            & \multicolumn{1}{c|}{\cellcolor[HTML]{FFFFFF}$C_2(x)$}            & \multicolumn{1}{c|}{\cellcolor[HTML]{FFFFFF}$C_2(y)$}            & \multicolumn{1}{c|}{\cellcolor[HTML]{FFFFFF}$M(y)$}            & \multicolumn{1}{c|}{\cellcolor[HTML]{FFFFFF}$M(x)$}            & \multicolumn{1}{c|}{\cellcolor[HTML]{FFFFFF}$M(xy)$}            & \multicolumn{1}{c|}{\cellcolor[HTML]{FFFFFF}-$3(z)$} & \multicolumn{1}{c|}{\cellcolor[HTML]{FFFFFF}-3-$(z)$} \\ \hline
$\mathbf{m} \parallel \hat{z}$ & $E$ & $I$ & $C_3$-$(z)$         & $C_3(z)$          & \multicolumn{1}{c|}{\cellcolor[HTML]{FFFFFF}$\Theta \cdot C_2(xy)$} & \multicolumn{1}{c|}{\cellcolor[HTML]{FFFFFF}$\Theta \cdot C_2(x)$} & \multicolumn{1}{c|}{\cellcolor[HTML]{FFFFFF}$\Theta \cdot C_2(y)$} & \multicolumn{1}{c|}{\cellcolor[HTML]{FFFFFF}$\Theta \cdot M(y)$} & \multicolumn{1}{c|}{\cellcolor[HTML]{FFFFFF}$\Theta \cdot M(x)$} & \multicolumn{1}{c|}{\cellcolor[HTML]{FFFFFF}$\Theta \cdot M(xy)$} & \multicolumn{1}{c|}{\cellcolor[HTML]{FFFFFF}-$3(z)$} & \multicolumn{1}{c|}{\cellcolor[HTML]{FFFFFF}-3-$(z)$} \\ \hline
$\mathbf{m} \parallel \hat{x}$ & $E$ & $I$ & $\Theta \cdot C_2(y)$ & $\Theta \cdot M(y)$ &                                                                   &                                                                  &                                                                  
&                                                                &                                                                &                                                                 &                                                      &                                                       
\\ \cline{1-5}
\end{tabular}
\caption*{Table S1. Point symmetries for different orientations of the magnetization. The xyz symbols refer to the primitive cell, e.g., $C_2(x)$ is a rotation of axis parallel to the first lattice vector, while the axis corresponding to $C_2(xy)$ is the sum of the first and second lattice vectors.}
\label{symmetries}
\end{table}

\begin{figure*}[h]\center
\includegraphics[width=9 cm,angle=0,keepaspectratio=true]{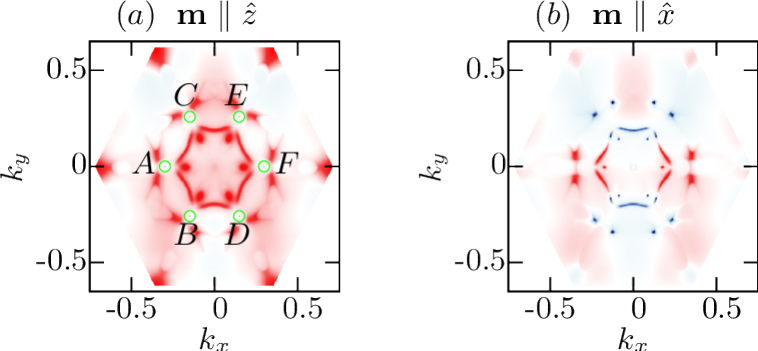}
	{\caption*{Figure S1. Flux of $\Omega_z$ integrated along $k_z$ as a function of $k_x$ and $k_y$ for magnetization parallel to $\hat{z}$ $(a)$ or $\hat{x}$ $(b)$ . Empty green circles correspond to the Weyl points.}}
\end{figure*}

\begin{figure*}[h]\center
\includegraphics[width=13 cm,angle=0,keepaspectratio=true]{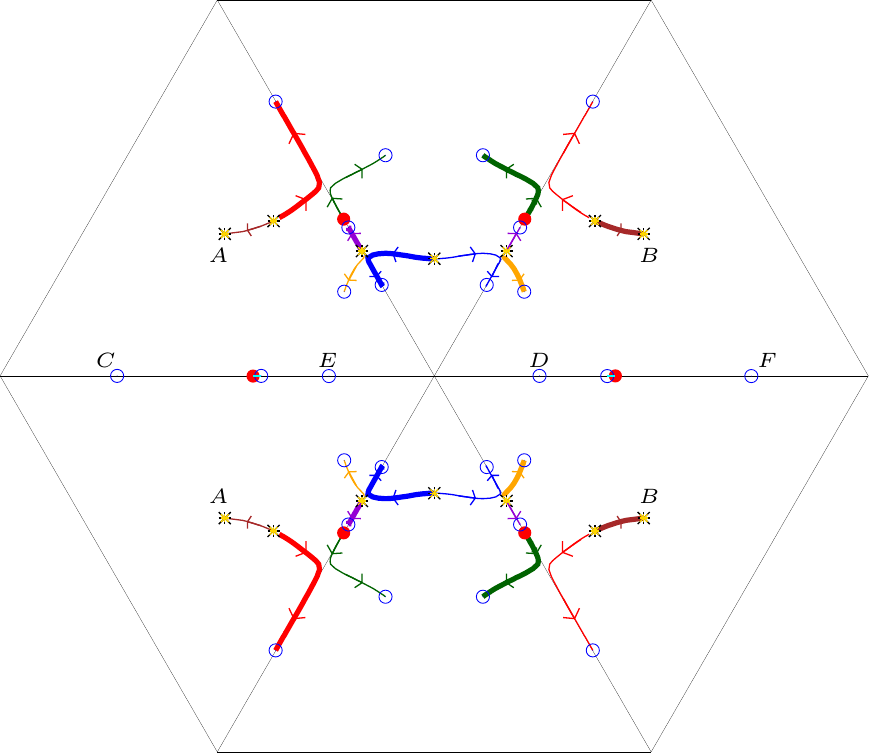}
	{\caption*{Figure S2. Projection of Weyl node trajectories on the plane $k_z=0$. The black lines are the projections onto this plane of the momentum-space primitive cell.
        Filled-red and empty-blue circles correspond to Weyl points at $\theta=0$ and $\theta=\pi/2$, respectively.
The thin (thick) lines indicate the trajectories of Weyl points of positive (negative) chirality and different line colors correspond to sets of Weyl nodes of different energy. The arrows indicate the sense of motion as $\theta$ increases.
	Weyl-node pairs ($A$,$B$), ($C$,$D$) and ($E$,$F$) annihilate each other by approaching the Brillouin zone border when their separation reaches a reciprocal lattice vector (the two latter pairs are stable only in a small range of angles close to $\theta=\pi/2$.}}
\end{figure*}

\clearpage
\subsection{Derivation of Eq. (3)}
In the presence of a gradient of temperature the current is \cite{PhysRevLett.97.026603} 
\begin{equation}
{\bf{j}} = -{\nabla T } \times \frac{e}{\hbar} \int d{\bf{k}} {\bf{\Omega}}({\bf{k}}) s(E_{{\bf{k}}}),
\label{eqj}
\end{equation}
where $d{\bf{k}} = \frac{d^3k}{(2\pi)^3}$ and the $s(E_{\bf{k}})$ is the entropy density of electrons with momentum $\bf{k}$:
\begin{equation}
s(E_{\mathbf{k}})=f_0(E_{\mathbf{k}})(E_{\mathbf{k}}-\mu)/T-k_B \text{log}\Big(1-f_0(E_{\mathbf{k}}) \Big)
\label{entropy}
\end{equation}
where $f_0$ is the equilibrium Fermi distribution. Notice that we include the $1/T$ factor in Eq. (\ref{entropy}) so that the entropy actually has the same units as the Boltzmann constant an that we omit the band index.
Assuming a gradient of temperature along $\hat{y}$, Eq. \ref{eqj} reduces to $j_x = -\alpha_{xy} \nabla T_y$ where $\alpha_{xy}$ is the Nernst conductivity. This tensor in general reads
\begin{equation}
	\alpha_{ab} = \frac{e}{\hbar} \int d\mathbf{k} \epsilon_{abc} \Omega_c(\mathbf{k}) s(E_{\mathbf{k}}).
\end{equation}
It is useful to write the volume element as $ d\mathbf{k} = \frac{1}{(2\pi)^3} d^2k_{\parallel} dk_{\perp}$, where $d^2k_{\parallel}$ and $dk_{\perp}$ are differential elements parallel and perdendicular to isoenergetic surfaces, respectively. Therefore, $dk_{\perp}= \frac{d\varepsilon} {|\nabla\varepsilon|}$ and $\alpha_{ab}$ reads
\begin{equation}
	\alpha_{ab} = \frac{e}{\hbar} \int \frac{d\varepsilon}{(2\pi)^3} s(\varepsilon) \iint d^2k_{\parallel} \epsilon_{abc} \frac{\Omega_c(\bf{k})} {|\nabla\varepsilon|},
\end{equation}
On the other hand, the anomalous Hall conductivity at zero temperature can be written as
\begin{equation}
	\sigma_{ab}(\varepsilon) = - \frac{e^2}{\hbar}\int\limits_{-\infty}^{\varepsilon} \frac{d\varepsilon'}{(2\pi)^3} \iint d^2k_{\parallel} \epsilon_{abc}\frac{\Omega_c(\bf{k})} {|\nabla\varepsilon|}. 
\end{equation}
Therefore, at zero temperature it follows that
\begin{equation}
	\frac{\partial\sigma_{ab}}{\partial \varepsilon} (\varepsilon) = -\frac{e^2}{\hbar} \frac{1}{(2\pi)^3} \iint\limits_{E_{\bf{k}}=\varepsilon} d^2k_{\parallel} \frac{\Omega_c(\bf{k})} {|\nabla\varepsilon|},
\end{equation}
where, as indicated, the double-integral is to be performed on the surface formed by states of energy equal to $\varepsilon$.
Notice that $\frac{\partial\sigma_{ab}}{\partial \varepsilon} (\varepsilon)$ is essentially a density of states weighted by the Berry curvature.
Therefore, the Nernst conductivity can be written as
\begin{equation}
	\alpha_{ab} = -\frac{1}{e} \int d\varepsilon \frac{\partial\sigma_{ab}}{\partial \varepsilon}(\varepsilon) s(\varepsilon) = \frac{1}{e} \int d\varepsilon \sigma_{ab}(\varepsilon) \frac{\partial s(\varepsilon)}{\partial \varepsilon}.
\end{equation}
The first equality is Eq. (3) in the main text.
In addition, from Eq. (\ref{entropy}) follows that $\frac{\partial s}{\partial \varepsilon} = \frac{(\varepsilon-\mu)}{T} \frac{\partial f}{\partial \varepsilon}$ and therefore
\begin{equation}
\alpha_{ab} = \frac{1}{e} \int d\varepsilon \sigma_{ab}(\varepsilon) \frac{\varepsilon-\mu}{T} \frac{\partial f}{\partial \varepsilon},
\label{alpha_form}
\end{equation}
which coincides with Eq. (8) of Ref. \cite{PhysRevLett.97.026603} (the different sign is compensated by the fact that here the derivative of $f$ is with respect to $\varepsilon$ instead of $\mu$).

The low temperature expansion can be obtained by defining $x = \beta(\varepsilon-\varepsilon_F)$, where $\varepsilon_F$ is the Fermi energy, and rewritting $\alpha_{ab}$ as:
\begin{align}
\alpha_{ab} &= -\frac{k_B}{e} \int dx \frac{e^x x}{(1+e^x)^2} \sigma_{ab}\Big(\frac{x}{\beta}+\varepsilon_F\Big) \nonumber \\
            &= -\frac{k_B}{e} \int dx \frac{e^x x}{(1+e^x)^2} \Big[\sigma_{ab}'(\varepsilon_F) \frac{x}{\beta} + \frac{1}{6}\sigma_{ab}'''(\varepsilon_F) \frac{x^3}{\beta^3} + ... \Big]
\end{align}
where we include only odd terms in the expansion because for the others the corresponding integrals vanish. Up to cubic term we obtain:
\begin{equation}
\alpha_{ab} = -\frac{\pi^2}{3}\frac{k_B^2 T}{e}  \Big[\sigma_{ab}'(\varepsilon_F) + \frac{7 \pi^2}{30} \sigma_{ab}'''(\varepsilon_F) k_B^2T^2 \Big]
\end{equation}
The first term is the usual Mott relation.

For the calculation of anomalous Nernst conductivity, we do not implement Eq. (2) of the main text but rather we use the Eq. (\ref{alpha_form}), 
which requires to compute the anomalous Hall conductivity as a function of energy, $\sigma_{ab}(\varepsilon)$ (see Ref. \cite{PhysRevLett.97.026603}).
In order to have a uniform convergence as the temperature is reduced, we find it convenient to compute $\sigma_{ab}(\varepsilon)$ on a logarithmic mesh centered at $\mu$, as shown in Figure S3 $(a)$. For the integraton, we used a mesh of $400^3$ $\mathbf{k}$-points in momentum space.
With this procedure the Nernst conductivity smoothly converges to the Mott relation at low temperatures, as shown in Figure S3$(b)$.

The results of entropy-flow density, $s_{f,ab}(\varepsilon) = \frac{\partial\sigma_{ab}}{\partial \varepsilon} s(\varepsilon)$, were obtained with a different calculation scheme aimed at improving the energy and momentum resolution close to the Fermi surface. 
Since the entropy flow involves the AHC only through derivatives with respect to energy, we set up a calculation of the AHC that only integrates the contribution from states in the energy windows [-20,20]\,meV. Namely, within this energy range, the derivative of the AHC with respect to energy is not affected by the contribution to the AHC from states away from the windows.
For this calculation, we first computed a mesh of $\mathbf{k}$-points having states in this energy windows and then we used this mesh to compute the AHC. The first step allows us to increase the mesh density from $400^3/V_{BZ}$ to $1500^3/V_{BZ}$ keeping a reasonable total number of points, where $V_{BZ}$ is the Brillouin zone volume.

\begin{figure}[h]\center
\includegraphics[width=9 cm,angle=0,keepaspectratio=true]{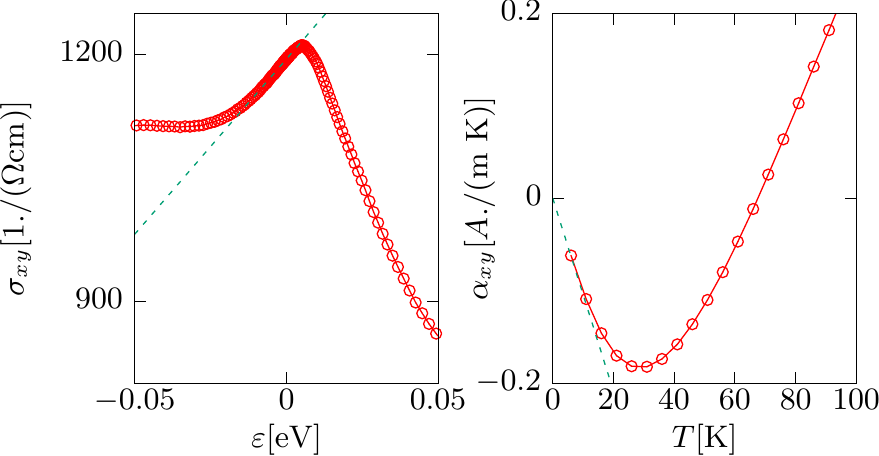}
	\caption*{Figure S3. \textit{Left:} AHC as a funtion of energy for magnetization parallel to $\mathbf{z}$. The dotted line at low energies is a linear fit.
	\textit{Right:} ANC as a function of temperature. The dotted line at low temperatures corresponds to the Mott relation.} 
\label{nernst_90}
\end{figure}

\twocolumngrid

\bibliography{ref}

\end{document}